# THE CHARGED BEAM DUMPS FOR THE INTERNATIONAL LINEAR COLLIDER

R. Appleby[#], Cockcroft Institute and the University of Manchester, UK
J.R.J. Bennett, T. Broome, C. Densham, CCLRC, UK
H. Vincke, CERN

*Abstract*

The baseline configuration of the International Linear Collider requires 2 beam dumps per interaction region, each rated to 18MW of beam power, together with additional beam dumps for tuning purposes and machine protection. The baseline design uses high pressure moving water dumps, first developed for the SLC and used in the TESLA design, although a gas based dump is also being considered. In this paper we discuss the progress made by the international community on both physics and engineering studies for the beam dumps.

## INTRODUCTION

A key aspect of the International Linear Collider (ILC) is the safe and controlled disposal of the high power electron, positron and photon beams produced by the machine. The work on a linear collider dump was started at the Stanford Linear Collider (SLC), where a small 2MW water dump was sufficient for the beam power [1], and continued by the TESLA project [2], where a larger water dump was proposed to handle the immense beam powers of the machine. The water dump has been adopted for the baseline of the ILC [3].

The baseline layout of the ILC consists of two interaction regions, which receive beam from the same two main linacs. This layout of the machine requires two full power beam dumps for each interaction region. Furthermore, the need to dump the beam at the end of the linac for commissioning and machine protection purposes requires two more beam dumps at the end of the linacs. These six charged beam dumps are required to be rated to 18MW for the 1 TeV machine.

The need to dump the beamstrahlung produced during the beam-beam interaction requires high power photon beam dumps. This beamstrahlung photon power can reach up to 3 MW. The beamstrahlung dump is part of the extraction line, and is common with the charged particle dump for the large crossing angle layout and separate for the small crossing angle layout.

There are also many smaller dumps and beam stops around the machine, which must form part of the beam dump baseline, and the need to consider the specialised dump required for photon-lepton and photon-photon collisions. The photon dump presents a unique challenge, as the uncharged photon beam cannot be rastered to reduce local power densities in the dump medium.

[#]r.b.appleby@dl.ac.uk

The ILC baseline dump layout was written at Snowmass 2005 by Appleby and Walz [3]. A pressurised water dump was specified as the baseline configuration, and the alternative was specified to be a noble gas based dump.

## BASELINE – WATER DUMP

The baseline is a pressurised water vortex dump, rated to 18 MW. This choice has many advantages: the dump has been studied in detail [1,2] and the problems of the dump are well known. The basic principle is to dump the energy of the beam in a region of cold, pressurised water. The water flow ensures the next part of the beam sees fresh water, and the outgoing heat is carried away to two heat exchanger loops. The solid backplate, placed well beyond the shower maximum, absorbs the residual part of the beam, and the water is separated from the vacuum of the extraction line by a thin window. The water pressure is 10bar, and the total volume of water in the system is 18 cubic metres. The dump length is 25m and is around 10-15m transversely.

The undisrupted beam spot size must be large enough to avoid window damage; this is achieved through rastering the beam and extraction line optics. Note the dump window should be actively cooled to avoid the thermal stress limit of the material. Volume boiling of the water is prevented by sufficient water flow rate and pressure. The dump shielding is critical to the dump design and neutron and muon studies are ongoing in this area.

## WATER DUMP PHYSICS STUDIES

The ongoing beam dump studies are focused on energy deposition, activation and shielding calculations. A study of pressure wave formation may also be required. The engineering studies are aimed at producing a dump design and full costing, according to the Global Design Effort (GDE) requirements and timescales. This work is beginning to happen; for example, the beam dumps is a new work package in the PPARC-funded LC-ABD, for the period 2007-2010, in the UK.

Current work is studying the energy deposition into the water dump. Figure 1 shows the energy density contours for a slice of the water dump when hit by a 500 GeV ILC electron beam. The beam dimensions are 3.83mm x 0.44mm. The aim is to tune the dump parameters to avoid volume boiling and ensure adequate water flow. The energy deposition at different radii in the water column is shown in figure 2 for a electron 500 GeV beam and in figure 3 for a 250 GeV electron beam.

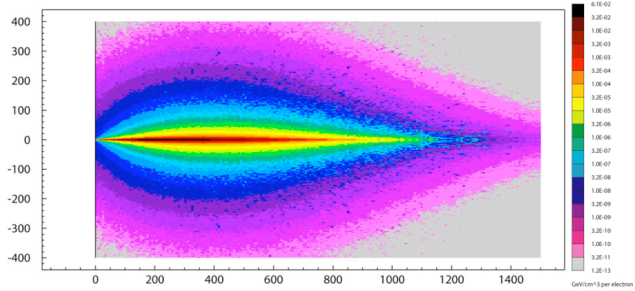

Figure 1: The energy density in the water dump from a 500 GeV electron beam with dimensions 3.83mm x 0.44mm. The horizontal (z) and vertical (radius) scales are in centimetres.

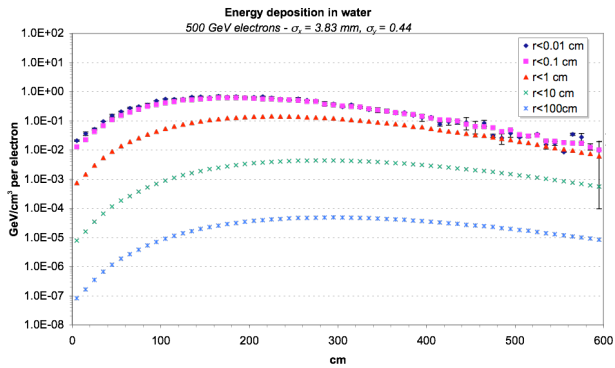

Figure 2: The energy deposition into the water dump, as a function of distance from the window, for varying radii inside the water column. The beam energy is 500 GeV.

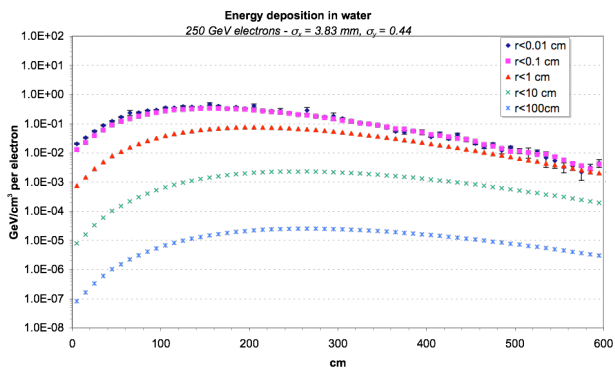

Figure 3: The energy deposition into the water dump, as a function of distance from the window, for varying radii inside the water column. The beam energy is 250 GeV.

## ALTERNATIVE – THE GAS DUMP

The alternative is a noble gas based dump [4]. The gas (argon looks the most promising) is contained in a thin column of 1km, which is enclosed in a water-cooled iron jacket. The gas acts as a scattering target for the beam, sending the beam energy transversely into the cooled iron. In essence, the gas dump acts as a passive beam expander.

The gas dump eases some problems of the water dump; for example the production of radioactive products is less than the water dump, although the design introduces some new issues. For example, heating and ionisation of the gas may be a problem. Furthermore, the dump maintenance may prove more problematic that the water dump because of the contained and activated gas.

Several dump combinations, or hybrid dumps, have been proposed. The combination of a gas dump, acting as a beam expander, in front of a reduced-complexity water dump, mitigates many problems of both the gas and water dump. This attractive idea is under study. A further possibility is the use of a gas dump in from of a solid material dump.

## GAS DUMP PRESSURE PROFILES

There is much work to do on the gas dump to bring it to the same level of development as the water dump. The work so far has looked the exploitation of the gas pressure profile along the length of the dump to tune the energy deposition. This pressure profile can be achieved using differential pumping and partitions. There is also the possibility of a windowless dump.

The energy deposition profile into argon of constant pressure gradient is the typical shower profile, with a maximum at some distance into the dump. If gas pumping is used to create a parabolic pressure gradient, the energy deposition profile can be tuned to uniform along the gas dump length. Such a profile would ease the dump design.

## CONCLUSIONS

The main charged beam dumps for the International Linear Collider present considerable technological challenge. We have presented physics studies of the water dump, and discussed the ongoing plans for a gas-based dump. The global physics and engineering work is beginning under the ILC global design effort.